\documentclass{llncs} 

 \usepackage{graphicx,color}
 \usepackage{amsmath,amssymb}
 \usepackage{wrapfig}
\usepackage{floatflt}
\usepackage{subfigure}
\usepackage{times}
\usepackage{multirow}  
\usepackage{verbatim}
\usepackage{enumitem}

\def\@thmcountersep{.}
\def\@thmcounterend{.}
\makeatother
\spnewtheorem{Example}[example]{Example}{\scshape}{}
\spnewtheorem{Definition}{Definition}[section]{\scshape}{}
\spnewtheorem{Theorem}[Definition]{Theorem}{\scshape}{}
\spnewtheorem{prop}[Definition]{Proposition}{\scshape}{}

\makeatletter
\renewcommand\paragraph{\@startsection{paragraph}{4}{\z@}%
                      {0.3\baselineskip}
                      {-0.5em \@plus -0.22em \@minus -0.1em}%
                      {\normalfont\normalsize\bfseries}}
\renewcommand\section{\@startsection{section}{1}{\z@}%
                      {-12\p@ \@plus -4\p@ \@minus -4\p@}%
                      {8\p@ \@plus 4\p@ \@minus 4\p@}%
                      {\normalfont\large\bfseries\boldmath
                       \rightskip=\z@ \@plus 8em\pretolerance=10000 }}
\renewcommand\subsection{\@startsection{subsection}{2}{\z@}%
                      {-12\p@ \@plus -4\p@ \@minus -4\p@}%
                      {8\p@ \@plus 4\p@ \@minus 4\p@}%
                      {\normalfont\bfseries\boldmath
                       \rightskip=\z@ \@plus 8em\pretolerance=10000 }}

\makeatother

\newcommand{\chv}{\ensuremath{\mathbf{v}}} 
\newcommand{\bfx}{\ensuremath{\mathbf{x}}} 
\newcommand{\bfp}{\ensuremath{\mathbf{p}}}

\newcommand{\bfc}{\ensuremath{\mathbf{\theta}}} 
\newcommand{\bfy}{\ensuremath{\mathbf{y}}} 

\newcommand{\bfX}{\ensuremath{\mathbf{X}}} 
\newcommand{\caL}{\ensuremath{\mathcal{L}}}

\newcommand{\bfO}{\ensuremath{\mathbf{O}}}
\newcommand{\bfo}{\ensuremath{\mathbf{o}}}
\newcommand{\bft}{\ensuremath{\mathbf{t}}}
\newcommand{\bfe}{\ensuremath{\mathbf{e}}} 
\newcommand{\bfs}{\ensuremath{\mathbf{s}}} 

\newcommand{\pr}[1]{\ensuremath{\mathit{Pr}\!\left(#1\right)}}

\begin{document}

\title{Optimal Observation Time Points in Stochastic Chemical Kinetics}
\author{
Charalampos Kyriakopoulos\inst{1}
\and Verena Wolf\inst{1}
}
\institute{
 Saarland University,
         Saarbr\"ucken, Germany}

 \maketitle
 
\begin{abstract}
 Wet-lab experiments, in which the dynamics within living cells are observed, are usually costly and 
 time consuming. This is particularly true if single-cell measurements are obtained using experimental techniques 
 such as flow-cytometry or fluorescence  microscopy. It is therefore important to optimize experiments
 with respect to the information  they provide about the system. In this paper we make a priori predictions of the amount of information that can be obtained from measurements. We focus on the case where the measurements are made to estimate parameters of a stochastic model of the underlying biochemical reactions. We propose a numerical scheme to approximate the Fisher information of future experiments at different observation time points and
 determine   optimal observation time points. 
 To  illustrate the usefulness of our approach, 
we apply our method to two interesting case studies.
 \end{abstract}

\section{Introduction}
The successful calibration of mathematical models of biological processes is usually   achieved by 
the interplay between computer simulations and wet-lab experiments. 
While both approaches are typically very time consuming,
wet-lab experiments are  costly compared to computer simulations, which is particularly true for 
 single-molecule techniques such as flow cytometry or fluorescence microscopy.
To keep the effort and costs as low as possible and to  employ wet-lab resources to have maximal gain,
 it is possible to run computer simulations before   measurements have been done in order to
 maximize the amount of information provided by the measurements. 
If the plan is to use the measurements for the estimation of unknown model parameters, then it is 
common practice to approach  the optimal experimental design problem by considering the Fisher information.
The Fisher information provides an approximation of the accuracy of parameter estimates
and if the information is maximal so is the (approximated) accuracy of the estimators.
While classical parameter estimation techniques compute the Fisher information of  parameter values estimated based on certain observations of the system, 
  it is also possible to do this computation before any observations are made. 
Thus, it is possible to compare the amount of information that a certain hypothetical experiment   provides
w.r.t. an unknown parameter.  

Here, we focus on stochastic models of biochemical reaction networks, for which the optimal experimental design problem has been addressed rarely in the past. We assume that  the kinetic constants of the chemical reactions have to be estimated and that flow cytometry measurements are possible. Our goal is to find the optimal
times at which measurements should be made to maximize the amount of information provided by the observations. Note that such observations are not correlated over time, thus, the results are independent 
single-cell measurements.  We do not optimize over other experiment design criteria such as 
the choice between the chemical species to observe. The reason is that the latter problem is, compared to
the optimal observation time points problem, much simpler since we only have to compare the Fisher
information for all possible combinations of observed species. 
Finding optimal observation times, however, is a challenging problem since the computation of the Fisher
information relies on a transient solution of the model. It  is therefore very costly and can usually not be done for a very large number of time points.
Thus, a sophisticated numerical procedure is necessary to efficiently determine optimal observation times.
Another problem is that the Fisher information depends not only on the time points the observations are made, but also on the unknown parameters value. 
We therefore assume that prior knowledge about the unknown parameters is available in the form of a prior
distribution. 
Given such a prior and a stochastic model with unknown kinetic constants, we determine those 
observation time points at which the expected Fisher information  is maximal.

There are two previous approaches to the optimal experimental design problem   for stochastic chemical kinetics. Both of them provide approximations of the Fisher information  and  focus on systems for which a direct numerical computation of the transient solution is too expensive. 
Komorowski et al. propose an approach that is based on the linear noise approximation, which assumes 
that molecules are present in sufficiently high copy numbers  \cite{komorowski}. 
However, many systems involve species present in small copy numbers leading to  
  highly skewed distributions \cite{ruess}. In such cases the  linear noise approximation yields poor
  approximations and the underlying master equation has to be solved directly.
Ruess et al. propose an approximation of the Fisher information  based on the  moments of the underlying probability distribution. Assuming that the number of observed cells is large,
they derive an  expression for the Fisher information that only involves moments
up to order four \cite{ruess}. This derivation is based on the 
  fact that sample mean and sample variance of the observations 
  are approximately normally distributed. 
  However, it turns out that  the  information provided by the sample mean and the sample variance do not suffice for the characterization of  skewed or bimodal distributions.  
In this paper we use a method based on the direct approximation of the underlying probability distribution 
since we found approximation errors up to 40\% when using the approach of Ruess et al.
We combine this approximation with a gradient-descent based optimization scheme in which
we, after sampling from the prior distribution, have to solve the underlying master equation only once
over time. The latter is achieved by a direct numerical approximation where the large state space of the model is truncated dynamically. This approach is a modification of an earlier method developed for the estimation
of parameters~\cite{CAV11}.

 After introducing the  stochastic model in Section~\ref{sec:cme},
 we give the mathematical background for the estimation of unknown parameters
 and explain how the   Fisher information and its time-derivatives are computed
 in Section~\ref{sec:fisherinf}.
In Section~\ref{sec:optimal} we present the optimal experiment design problem and
propose a numerical method for finding optimal observation times.
Finally, we report on experimental results for two reaction networks 
(Section~\ref{sec:exp})
and conclude with a discussion of the results in Section~\ref{sec:conc}.

 \section{Discrete-state Stochastic Model}\label{sec:cme}
  According to Gillespie's theory of stochastic chemical kinetics, 
 a well-stirred mixture of $n$ molecular species in a volume 
 with fixed size and fixed temperature can be represented as 
 a continuous-time Markov chain 
 $\left\{\bfX_t,t\ge 0\right\}$~\cite{gillespie77}. 
 The random vector $\bfX_t$ 
 describes the chemical populations at time $t$, i.e., the $i$-th entry  is 
 the   number of molecules of type $i\in\{1,\ldots,n\}$ at time 
$t$.
 Thus, the state space of $\bfX$ is $\mathbb Z^n_+=\{0,1,\ldots\}^n$.
The state changes of $\bfX$ are triggered by the occurrences of 
chemical reactions, which are of $m$ different types.  
For $j\in\{1,\ldots,m\}$  
let the row vector $\chv_j\in\mathbb Z^n$ be the nonzero
\emph{change vector} of the $j$-th reaction type, that is, 
$\chv_j=\chv_j^-+\chv_j^+$ where $\chv_j^-$ contains only 
non-positive entries, which specify how many molecules of each
species are consumed (\emph{reactants}) if an instance of the 
reaction occurs. 
The vector  $\chv_j^+$ contains only non-negative
 entries, which  specify how many molecules of each
species are produced (\emph{products}). 
Thus, if $\bfX_t=\bfx$ for some $\bfx \in \mathbb Z^n_+$ with 
$\bfx+\chv_j^-$ being non-negative, then $\bfX_{t+dt}=\bfx+\chv_j$
is the state of the system after the occurrence of the $j$-th 
reaction within the infinitesimal time interval $[t,t+dt)$.
 
Each reaction 
type has an associated \emph{propensity function}, denoted by 
$\alpha_1,\ldots,\alpha_m$, which is such that 
$\alpha_j(\bfx)\cdot dt$ is the probability that, given $\bfX_t=\bfx$, 
one instance of the $j$-th reaction occurs within $[t,t+dt)$.
Often the value $\alpha_j(\bfx)$ is chosen 
proportional to the number of distinct
reactant combinations in state $\bfx$, known as the law of
mass action.
However, for many reactions the proportionality constant $\theta_j$ is  
 unknown and has to be estimated based on measurements.
For instance, if we have two (distinct) reactants (i.e. $ \chv_j^-=-\bfe_i-\bfe_\ell$) then 
$\alpha_j(\bfx)=\theta_j \cdot x_i\cdot x_\ell$
where $x_i$ and $x_\ell$ are the corresponding entries of $\bfx$, $i\neq \ell$, $\theta_j>0$, and 
$\bfe_i$ is the vector with the $i$-th entry $1$ and all other
entries $0$. In the sequel we do not restrict the form of $\alpha_j$ 
but only assume that its derivative w.r.t. some unknown parameter $\theta_j$ exists. Sometimes we will make the dependence of $\alpha_j$ on $\theta_j$ explicit by writing  $\alpha_j(\bfx,\theta_j)$ instead of  $\alpha_j(\bfx)$.

\begin{example}
 \label{ex:crystallization}
We consider a simple crystallization process
that involves four chemical species, namely A, B, C and D. Thus, the entries of the random vector 
$\bfX_t$ give the numbers of molecules of types A, B, C and D at time $t$. The two possible
reactions are 
2A  $\to$ B and A $+$ C $ \to$ D. 
Thus, $\chv_1 = (-2, 1, 0, 0)$, $\chv_2 = (-1, 0, -1, 1)$. 
For a state $\bfx = (x_A, x_B, x_C, x_D)$, the propensity functions are
$\alpha_1(\bfx) = \theta_1 \cdot \binom{x_A}{2}$ and $\alpha_2(\bfx) = \theta_2 \cdot x_A \cdot x_C$. 
Note that given an initial state $\bfx_0$ the set of reachable states is a finite subset of $\mathbb N^4$.   
 \end{example} 

In general, the reaction rate constants $\theta_j$ refer  to the 
probability that a randomly selected 
pair of reactants collides and undergoes the $j$-th chemical reaction. 
It depends on the volume and the temperature of the system as well
as on the microphysical properties of the reactant species. 
Since reactions of higher order (requiring 
more than two reactants) are usually the result of several successive lower 
order reactions, we do not consider the case of more than two reactants.


\textbf{The Chemical Master Equation.}
 For $\bfx\in \mathbb Z^n_+$ and $t\ge 0$, let $p_t(\bfx)$ denote the 
probability $\pr{\bfX_t=\bfx}$. Given $\chv_1^-,\ldots, \chv_m^-$, $\chv_1^+,\ldots, \chv_m^+$, 
$\alpha_1,\ldots,\alpha_m$, and some initial distribution $\bfp_0$, the  
Markov chain $\bfX$ is uniquely specified
and its  evolution  is given by the 
 chemical master equation (CME)
\begin{equation}\label{eq:CMEstate}
\textstyle \frac{d}{dt}p_t(\bfx)=\displaystyle\sum_{j:\bfx-\chv_j^-\ge 0} \alpha_j(\bfx-\chv_j) 
p_t(\bfx-\chv_j)-\alpha_j(\bfx)
p_t(\bfx).
\end{equation}
If we use $\bfp_t$ to denote the row vector 
with entries $p_t(\bfx)$, then the vector form of the CME becomes
 \begin{equation}
 \label{eq:CME}
 \begin{array}{r@{\ }c@{\ }l}
 \frac{d}{dt} \bfp_t &=&  \bfp_tQ,
  \end{array}
 \end{equation}
where  $Q$ is the infinitesimal generator matrix of $\bfX$
  with $Q(\bfx,\bfy)=\alpha_j(\bfx)$ if $\bfy=\bfx+\chv_j$
and $\bfx+\chv_j^-\ge 0$.
Note that, in order to simplify our presentation, we assume here that all vectors $\chv_j$
are distinct. All remaining entries of $Q$ are zero except for the diagonal 
entries which are equal to the negative row sum. 
The ordinary first-order differential equation in~\eqref{eq:CME}  is a direct
consequence of the Kolmogorov forward equation.
Since $\bfX$ is a regular Markov process,~\eqref{eq:CME} 
has the general solution 
$
  {\bfp}_t={\bfp}_0\cdot e^{Qt},
$
 where $e^{A}$ is the matrix exponential of a matrix $A$.
If the state space of $X$ is infinite, 
then we can only compute approximations of  ${\bfp_t}$.
But even if $Q$ is finite, its size is often large 
because it grows exponentially with the number of state variables. 
Therefore standard numerical solution techniques 
for  systems of  first-order linear equations of the form of~\eqref{eq:CME}
are infeasible.
The reason is that the number  of nonzero entries in $Q$ often 
exceeds the available 
 memory capacity  for systems of realistic size. 
 If  the populations of all species remain
 small (at most a few hundreds) 
 then  the CME can be efficiently
  approximated using projection methods~\cite{sliding,Munsky06,krylovSidje}
 or fast uniformization methods~\cite{FAUIET,Inexact}.
The idea of these methods is to avoid an exhaustive state space 
exploration and, depending on a certain time interval, restrict the analysis
of the system to a subset of states. 

Here, we are also interested in the partial
derivatives of $\bfp_t$ w.r.t. the 
reaction rate constants $\bfc=(\theta_1,\ldots,\theta_m)$. 
In the sequel we will write $\bfp_t(\bfc)$ instead of $\bfp_t$ to make 
the dependency on $\bfc$ explicit and the entry of $\bfp_t(\bfc)$ that 
corresponds to state $\bfx$ will be
denoted by $p_t(\bfx;\bfc)$.
Moreover, we define the  row vectors $\bfs^j_t(\bfc)$  as the
derivative of $\bfp_t(\bfc)$ w.r.t. $\theta_j$, i.e.,
$$
\textstyle\bfs_t^j(\bfc)=\frac{\partial \bfp_t(\bfc)}{\partial \theta_j} =
\lim_{\Delta h\to 0}\frac{\bfp_t(\bfc+{\bf \Delta
h}^{(j)})-\bfp_t(\bfc)}{\Delta h},
$$
where the vector ${\bf \Delta h}^{(j)}$ is zero everywhere except for the $j$-th
position that is equal to $\Delta h$. We denote the entry in $\bfs_t^j(\bfc)$
that corresponds to state $\bfx$ by $s_t^j(\bfx,\bfc)$.
Derivating ~\eqref{eq:CME}, we find that $\bfs_t^j(\bfc)$ is the unique solution of 
the following linear system of ODEs 
 \begin{equation}
 \label{eq:derivCME}
 \begin{array}{r@{\ }c@{\ }l}
 \frac{d}{dt} \bfs_t^j(\bfc) &=&  \bfs_t^j(\bfc)Q +
\bfp_t(\bfc)\frac{\partial}{\partial \theta_j}Q,
  \end{array}
 \end{equation}
where $j\in\{1,\ldots, m\}$.
The initial condition is $s_0^j(\bfx,\bfc)=0$ for all $\bfx$ and $\bfc$ 
since $p_0(\bfx;\bfc)$ is independent of $\theta_j$.  

 \section{Observations and Fisher Information}\label{sec:fisherinf}
Following the notation in~\cite{Timmer}, we assume that observations
of a biochemical network are made at time instances 
$t_1,\ldots,t_R\in\mathbb R_{\ge 0},$
where $t_1 \leq \ldots \leq t_R$.
The entries of the random vector $\bfO_{t_k}\in\mathbb R^n$ describe the 
molecule numbers observed at time 
$t_k$ for  $k\in\{1,\ldots,R\}$. 
Since these observations are typically subject to measurement errors,
we may assume that $\bfO_{t_k} = \bfX_{t_k} + \epsilon_{t_k},$
where the entries of the error terms $\epsilon_{t_k}$ are independent and 
identically normally 
distributed with mean zero and standard deviation $\sigma$. Note that 
$\bfX_{t_k}$ is the true population vector
 at time $t_k$.
Clearly, this implies that, conditional on $\bfX_{t_k}$,
the random vector $\bfO_{t_k}$ is independent of all 
other observations as well as  independent of the history of $\bfX$ before 
time $t_k$.

Let $f$ denote the joint density of $\bfO_{t_1}, \ldots, \bfO_{t_R}$ and assume
that $\bfo_{1}, \ldots, \bfo_{R}\in \mathbb R^n$.
Then the likelihood of the observation sequence $\bfo_{t_1}, \ldots, \bfo_{t_R}$ is
\begin{equation}\label{eq:likelihood}
\begin{array}{lcl}
\caL&=&f\left(\bfO_{t_1}=\bfo_{1},\ldots,\bfO_{t_R}=\bfo_{R}\right)\\[1ex]
&=&\sum_{\bfx_1}\ldots \sum_{\bfx_R} f\left(\bfO_{t_1}=\bfo_{1}, \ldots, \bfO_{t_R}=\bfo_{R}
\mid \bfX_{t_1}=\bfx_1,\ldots,\bfX_{t_R}=\bfx_R\right)\\[1ex]
&&   \pr{\bfX_{t_1}=\bfx_1, \ldots, \bfX_{t_R}=\bfx_R}.
\end{array}
\end{equation}
We assume that for the unobserved process $\bfX$
we do not know the values of the rate constants 
$\theta=(\theta_1,\ldots,\theta_m)$ and  our aim is to estimate these 
constants.

In the sequel, in order to keep the notation simpler, we
assume no measurement errors in our observations. 
Nevertheless, it is straightforward to extend our proposed optimal design procedure 
to the case where the measurements are not exact. As shown in~\cite{CAV11} 
this only introduces additional weights during  the calculation of the likelihood. 
In this case, also, one can consider the standard deviation of
the error terms,  $\sigma$, as an unknown parameter and apply the proposed design.
Here, though, we assume that $\bfO_{t_k} = \bfX_{t_k}$ for all $k$ and 
for a concrete observation sequence $\bfx_{1}, \ldots, \bfx_{R}\in \mathbb N^n$ the likelihood becomes
\begin{equation}\label{eq:likelihood_here}
\begin{array}{lcl}
\caL = \pr{\bfX_{t_1}=\bfx_1,\ldots,\bfX_{t_R}=\bfx_R}.
\end{array}
\end{equation}
 Of course,  $\caL$ depends 
on the chosen rate parameters $\bfc$ since the probability 
measure $ \pr{\cdot }$ does.  
 When necessary, we will make this 
dependence explicit by writing $\caL(\bfc)$ instead of $\caL$.
Since our observations are equal to the true state $\bfX_{t_k}$
at time $t_k$, we write $\caL(\bfX;\bfc)$ where with some abuse of notation $\bfX$ now refers
to the sequence $\bfX_{t_1}, \ldots, \bfX_{t_R}$. 
We now seek constants $\bfc^*$   such that 
\begin{equation}\label{eq:MLEestimator}
 { \bfc^* = \mathrm{argmax}_{\bfc} \;\caL(\bfX; \bfc)},
\end{equation}
where  the maximum is taken over all   vectors
$\bfc$ with all components strictly positive. 
This optimization problem is known as the maximum likelihood 
problem~\cite{citeulike:821121}.
Note that $\bfc^*$ is a random variable, in the sense it
depends on the (random) observations $\bfX=(\bfX_{t_1}, \ldots, \bfX_{t_R})$.

The maximum likelihood estimator $\bfc^*$ is known to be asymptotically normally distributed
and its covariance matrix approaches the Cram\'er-Rao bound $  {\cal{I}}_\bfX(\theta)^{-1},$ where 
${\cal{I}}_\bfX(\theta)$ is the Fisher information matrix (FIM). Note that this bound is a lower bound
on the estimator's covariance matrix. 
It is commonly used to derive confidence intervals for the estimated values of the parameters.
Previous experimental results show that the variances approximated based on the FIM are close to the
variances approximated based on many repetitions of the experiments and the estimation procedure~\cite{CAV11}. 
Thus, in order to have an accurate estimation of the parameters the (co-)variances should be as
small as possible, i.e., $  {\cal{I}}_\bfX(\theta)$ should be large to achieve tight confidence intervals
for $\bfc^*$.

Given an observation sequence,  $\bfX$ of the process, the entry of the FIM that corresponds to 
  the unknown parameters $\theta_i$ and $\theta_j$, $1\leq i, j \leq m$   
 is defined as
\begin{equation}\label{eq:FIM_1}
 \begin{array}{lcl}
\big( {\cal{I}}_\bfX(\theta) \big)_{i, j} 
&=& \mathbb{E}_\bfX\Big[\Big(\frac{\partial}{\partial \theta_i} \mathrm{log}\;\mathcal{L}(\bfX; \theta) \Big) \Big(\frac{\partial}{\partial \theta_j} \mathrm{log}\;\mathcal{L}(\bfX; \theta) \Big)\Big].\\[1ex]
\end{array}
\end{equation} 
Note that the expectation is taken w.r.t. the observation sequence $\bfX=(\bfX_{t_1}, \ldots, \bfX_{t_R})$.
 Under certain (mild) regularity conditions that hold for the likelihoods we consider here, Eq.~\eqref{eq:FIM_1} can be equivalently written as~\cite{vanBos2007parameter}
\begin{equation}\label{eq:FIM_2}
 \begin{array}{lcl}
\big( {\cal{I}}_\bfX(\theta) \big)_{i, j} 
&=& - \mathbb{E}_\bfX\Big[\frac{\partial^2}{\partial \theta_i \partial \theta_j} \mathrm{log}\;\mathcal{L}(\bfX; \theta)\Big]\\[1ex]
&=& - \sum_{\bfx_1,\ldots,\bfx_R} \pr{\bfX_{t_1} = \bfx_1, \ldots, \bfX_{t_R} = \bfx_R; \theta}\\[1ex]
 & & 
\quad  \quad \frac{\partial^2}{\partial \theta_i \partial \theta_j} \mathrm{log} \pr{\bfX_{t_1} = \bfx_1, \ldots, \bfX_{t_R} = \bfx_R; \theta}. 
\end{array}
\end{equation} 
Note that if  the observations $\bfX_{t_1}, \ldots, \bfX_{t_R}$ are  independent observations 
 then the $(i, j)$th entry of the FIM is such that  
 \begin{equation}\label{eq:FIM_indep}
 \begin{array}{lcl}
\big( {\cal{I}}_\bfX(\theta) \big)_{i, j} 
&=& \displaystyle \sum_{k=1}^R \big( {\cal{I}}_{\bfX_{t_k}}(\theta) \big)_{i, j}
\end{array}
\end{equation}
where ${\cal{I}}_{\bfX_{t_k}}(\theta) $ is the Fisher information 
matrix of a single observation 
$\bfX_{t_k}$ at time $t_k$.
The above means that the information of a sequence of $R$ independent
observations is simply the sum of the information of each. This makes the computation of the total information
  easier   than in the dependent case since it is enough   to solve the CME along with the partial derivatives
$\bfs_t^j(\bfc)$, for all $j$, until time point $t_R.$ This can be easily seen by exploiting \eqref{eq:FIM_1} for $\bfX_{t_k}$.
\begin{equation}\label{eq:FIM_exploit}
 \begin{array}{lcl}
\big( {\cal{I}}_{\bfX_{t_k}}(\theta) \big)_{i, j} 
&=& \mathbb{E}_{\bfX_{t_k}}\Big[\Big(\frac{\partial}{\partial \theta_i} \mathrm{log}\; \pr{\bfX_{t_k}; \theta}\Big) \Big(\frac{\partial}{\partial \theta_j} \mathrm{log}\;\pr{\bfX_{t_k}; \theta} \Big)\Big]\\[1.5ex]
&=&  \displaystyle \sum_{\bfx_k} \Big(\frac{\partial}{\partial \theta_i} \mathrm{log}\; p_{t_k}(\bfx_k; \theta) \Big) \Big(\frac{\partial}{\partial \theta_j} \mathrm{log}\; p_{t_k}(\bfx_k; \theta) \Big)  p_{t_k}(\bfx_k; \theta) \\[1.5ex]
&=&  \displaystyle\sum_{\bfx_k}  \frac{ 1}{p_{t_k}(\bfx_k; \theta)}\frac{\partial}{\partial \theta_i} p_{t_k}(\bfx_k; \theta) \; \frac{\partial}{\partial \theta_j} p_{t_k}(\bfx_k; \theta) ,
\end{array}
\end{equation} 
where the sums run over all possible states $\bfx_k$ that can be observed at time $t_k$.
Using the notation of the previous section, we get
\begin{equation}\label{eq:FIM_exploit_final}
 \begin{array}{lcl}
\big( {\cal{I}}_{\bfX_{t_k}}(\theta) \big)_{i, j} 
&=& \displaystyle \sum_{\bfx_k}  \frac{ s_{t_k}^i(\bfx_k; \theta) \; s_{t_k}^j(\bfx_k; \theta) }{p_{t_k}(\bfx_k; \theta)}.
\end{array}
\end{equation}
Derivating \eqref{eq:FIM_exploit_final} we can also compute the derivative of the FIM w.r.t. the time point $t_k$. 
\begin{equation}\label{eq:FIM_der}
 \begin{array}{lcl}
 &&\mbox{}\\
\frac{\partial}{\partial t} \big({\cal{I}}_{\bfX_{t_k}}(\theta) \big)_{i, j} 
&=& \displaystyle \sum_{\bfx_k}  \frac{  \frac{\partial}{\partial t} s_{t_k}^i(\bfx_k; \theta) \;  s_{t_k}^j(\bfx_k; \theta) + s_{t_k}^i(\bfx_k; \theta) \; 
\frac{\partial}{\partial t}s_{t_k}^j(\bfx_k; \theta) } {p_{t_k}(\bfx_k; \theta) } \\[1ex] 
&&  - \displaystyle \sum_{\bfx_k} \frac{s_{t_k}^i(\bfx_k; \theta) \;  s_{t_k}^j(\bfx_k; \theta) \; \frac{\partial}{\partial t} p_{t_k}(\bfx_k; \theta)}
{p_{t_k}(\bfx_k; \theta)^2}.
\end{array}
\end{equation}
The time derivative of the FIM is particularly useful for an efficient gradient-based optimization scheme to find the time points that provide the 
maximum information which is the main goal of the next section.

\section{Optimal Observation Time Points}
\label{sec:optimal}
Assume now that an experiment is planned where the system is observed at time points $t_1, \ldots, t_R$
and that the observations will be independent  (e.g. if the chosen measurement technique is flow cytometry).
We want to find the optimal time points to take our observations, i.e. those which yield the maximum information
for the system parameters $\theta$.
However, the intrinsic problem in experiment design is that the parameter values $\theta$ are, of course, unknown before the experiment is set up. Here, the chosen approach
to overcome this obstacle is to search for the observation time points that maximize the determinant of the \emph{expected} FIM
when one assumes a prior distribution for the unknown parameters as suggested, for instance, in \cite{merle}. In other words our goal is to find 

\begin{equation}\label{eq:optimization}
 \begin{array}{lcl}
\bft^* = \underset{\bft = (t_1, \ldots, t_R)}{\mathrm{argmax}}\; \mathrm{det}\big( \mathbb{E}_\theta[\mathcal{I}(\bft, \theta)] \big).
\end{array}
\end{equation}
This criterion is known to be robust because it incorporates a prior belief for $\theta$.
%
%
Note that in the sequel we write $\mathcal{I}(\bft, \theta)$ instead of ${\cal{I}}_{\bfX}(\theta) $ to explicitly 
indicate that the information is a function of the sequence of time points, $\bft$, at which the observations are to be taken. Also, remember that in the previous section we showed that $\mathcal{I}(\bft, \theta) =  \sum^R_{k = 1}\mathcal{I}(t_k, \theta),$ when we assume that the observations are independent.

It is worth mentioning that sometimes not all reaction rate constants are of interest, in which case we partition the parameter vector $\theta = [\theta_{(1)}, \theta_{(2)}]$ in parameters of interest $\theta_{(1)}$ and nuisance parameters $\theta_{(2)}$. Then, for a specific time point $t$, ${\cal{I}}(t, \theta)$ is replaced by the matrix ${\cal{I}}_s(t, \theta)$
which is defined as
\begin{equation}\label{eq:I_s}
 \begin{array}{lcl}
{\cal{I}}_s(t, \theta) 
&=& {\cal{I}}_{11}(t, \theta) - {\cal{I}}_{12}(t, \theta)\; {\cal{I}}_{22}^{\;-1}(t, \theta) \;{\cal{I}}_{12}^{\;\intercal}(t, \theta), \mathrm{where}\\[1ex]
{\cal{I}}(t, \theta)
&=& \left[ \begin{array}{cc}
{\cal{I}}_{11}(t, \theta) &  {\cal{I}}_{12}(t,\theta)\\
{\cal{I}}_{12}^{\;\intercal}(t, \theta)  & {\cal{I}}_{22}^{\;-1}(t, \theta)\\
\end{array} \right].
\end{array}
\end{equation}
Here, subscript $s$ indicates that we only consider derivatives of the parameters of interest, i.e.,
the matrices ${\cal{I}}_{11}(t, \theta)$, ${\cal{I}}_{22}(t, \theta)$ contain information about the variance 
of $\theta_{(1)}$ and $\theta_{(2)}$, respectively, while 
${\cal{I}}_{12}(t, \theta)$ approximates  the covariance matrix of $\theta_{(1)}$ and $\theta_{(2)}.$ 
Hence,   given a prior of $\theta$ and having computed the matrices
$ {\cal{I}}(t_k, \theta) $  for all $\theta$ and all $k$,
it is straightforward to compute the matrix $\mathbb{E}_\theta[\mathcal{I}_s(\bft, \theta)] $ and then $\mathrm{det}\big( \mathbb{E}_\theta[\mathcal{I}_s(\bft, \theta)] \big)$. Similarly, if we have, in addition, $\frac{\partial}{\partial t} {\cal{I}}(t, \theta)$  for all $\theta$ and all $t\in\{t_1,\ldots,t_R\}$ we can also compute $\frac{\partial}{\partial t} \mathrm{det}\big( \mathbb{E}_\theta[\mathcal{I}_s(\bft, \theta)] \big)$
as follows. 
From Jacobi's formula it holds that for any square matrix $A$
\begin{equation}\label{eq:dt_det_I_s}
 \begin{array}{lcl}
\frac{\partial}{\partial t} \mathrm{det}(A)
&=& \mathrm{tr}(\mathrm{adj}(A))\frac{\partial}{\partial t} A, 
\end{array}
\end{equation}
where $\mathrm{tr}(A)$ is defined as the sum of the elements of the main diagonal of $A$ and $\mathrm{adj}(A)$
is the adjoint matrix of $A$. Here, $A = \mathbb{E}_\theta[\mathcal{I}_s(\bft, \theta)]$ and, derivating Eq.~\eqref{eq:I_s}, we can compute 
$\frac{\partial}{\partial t}\mathbb{E}_\theta[\mathcal{I}_s(\bft, \theta)]$ by
exploiting  known matrix calculus identities.
\begin{equation}\label{eq:dt_det_I_s}
 \begin{array}{lcl}
\frac{\partial}{\partial t} \mathbb{E}_\theta [{\cal{I}}_s(\bft, \theta) ]
&=& \mathbb{E}_\theta  [\frac{\partial}{\partial t} {\cal{I}}_{11}(\bft, \theta) -  \frac{\partial}{\partial t} {\cal{I}}_{12}(\bft, \theta)\;{\cal{I}}_{22}^{\;-1}(\bft, \theta)\; {\cal{I}}_{12}^{\;\intercal}(\bft, \theta) \\[1ex]
&& -  {\cal{I}}_{12}(\bft, \theta)\; \frac{\partial}{\partial t} {\cal{I}}_{22}^{\;-1}(\theta)\; {\cal{I}}_{12}^{\;\intercal}(\bft, \theta) \\[1ex]
&& -  {\cal{I}}_{12}(\bft, \theta)\; {\cal{I}}_{22}^{\;-1}(\bft, \theta)\; \frac{\partial}{\partial t} {\cal{I}}_{12}^{\;\intercal}(\bft, \theta)],
\end{array}
\end{equation} where  the derivative of the inverse is computed as 
\begin{center}
$\frac{\partial}{\partial t} {\cal{I}}_{22}^{\;-1}(\bft, \theta) = -{\cal{I}}_{22}^{\;-1}(\bft, \theta) \; \frac{\partial}{\partial t} {\cal{I}}_{22}(\bft, \theta) \; {\cal{I}}_{22}^{\;-1}(\bft, \theta).$
\end{center}
The main computational effort in the search for the   optimal experiment  is that 
for finding $\bft^*$ as defined in Eq.~\eqref{eq:optimization}, we have to solve the CME after sampling from the
prior of $\theta$. It is generally not possible to find $\bft^*$ from a single solution of the CME since we need
to average over all possible values for $\theta=(\theta_1,\ldots,\theta_m)$.
Therefore, we propose the following gradient descent based 
procedure to approximate local maxima of the determinant of the expected FIM:
 \begin{enumerate}
\setlength{\itemsep}{10pt}
\item Given a prior $\mu_\theta$ for the distribution  of $\bfc$, 
 sample values
$ \theta^{(1)},\ldots, \theta^{(N)} \thicksim \mu_\theta.$ 
\item Choose a sequence $\bft_{\mathrm{next}} = (t_1, \ldots, t_R)$ of time points 
and for each sample of $\theta^{(i)}$ compute $\mathcal{I}_s(\bft_{\mathrm{next}}, \theta^{(i)})$ and $\frac{\partial}{\partial t}\mathcal{I}_s(\bft_{\mathrm{next}}, \theta^{(i)})$ for $t\in\{t_1,\ldots,t_R\}$.
 \item 
Return approximations of $\mathrm{det}\big(\mathbb{E}_\theta[\mathcal{I}_s(\bft, \theta)]\big)$
and $\frac{\partial}{\partial t} \mathrm{det} \big(\mathbb{E}_\theta[ \mathcal{I}_s(\bft, \theta)]\big)$ by averaging over the  results for $i=1,\ldots,N$. 
\item Following the gradient choose $\bft_{\mathrm{next}} = (t'_1, \ldots, t'_R)$ and repeat from 2 until you find a local maximum.
\end{enumerate}
Clearly, an approximation of the global maximum is found by starting the local gradient based search from multiple initial points. The initial points used in the second step can be chosen randomly or according to some heuristics as
it is usual for global optimization methods. 
A technical but computationally important detail is that there is no need to solve the CME for every sequence of time points  that is considered in the optimization algorithm. 
We can solve the CME once and keep the values of the Fisher information matrix and its derivatives over time  and recall it for every new sequence of time points (up to a chosen discretization). 
Consequently, the total computational effort of the above optimization procedure is to solve once the CME until time point equal to the maximal value of $t_R$ encountered during the optimization.

Certainly, the above optimality criterion is by no means the only possible choice. 
A slightly different approach, for instance, would be to maximize the expected determinant of FIM \cite{pronzato1}.
Alternatively, if no prior is available, it is also possible to consider $\max_\bft \min_\theta \mathrm{det}\big( \mathcal{I}_s(\bft, \theta)\big)$ to make sure that for any choice of $\theta$ the experiment provides maximal information.
Here, we assume that a prior is available because, most probably,
one experiment has been already done in order to acquire some prior knowledge for the parameters. At last,  an additional advantage of a robust optimal design is that in case there is the option to perform more than two experiments in total, 
the above procedure can be used in iterations alternating
between experiments and the update of the prior of $\theta$ 
via parameter estimation from the real observations.

\section{Experimental Results}
\label{sec:exp}
We consider two biochemical reaction networks to which we apply  our experiment design procedure, namely the crystallization model, described in Example \ref{ex:crystallization}, 
and the so-called exclusive switch model~\cite{loinger-lipshtat-balaban-biham07}. 
The crystallization model is a very simple example because it has a finite state space and
 the CME can be integrated directly if the initial molecule numbers are not particularly large. 
However, for large initial conditions a transient solution is only possible if the state space
is dynamically truncated as suggested in \cite{FAUIET} (see also \cite{MLEJournal}).
The second example is infinite in two dimensions and its distribution is bimodal. 
Thus, for this system we solved the CME using a dynamic truncation of the state space whenever necessary.
We chose the truncation threshold $e^{-20}$ and the dynamic truncation of 
the states are based on the ratio in  Eq.~\eqref{eq:FIM_exploit_final}, rather than only 
on the value of $p_{t}(\bfx; \theta)$. More precisely, a state $\bfx$ is considered as significant at time $t$
whenever this ratio is greater than the truncation threshold while in  \cite{FAUIET}  it is 
only the current probability that determines whether a state is considered or not at time $t$.

\subsection{Crystallization}
\begin{floatingfigure}[l]{0.5\textwidth}
\begin{center}
 \includegraphics[width=0.5\textwidth]{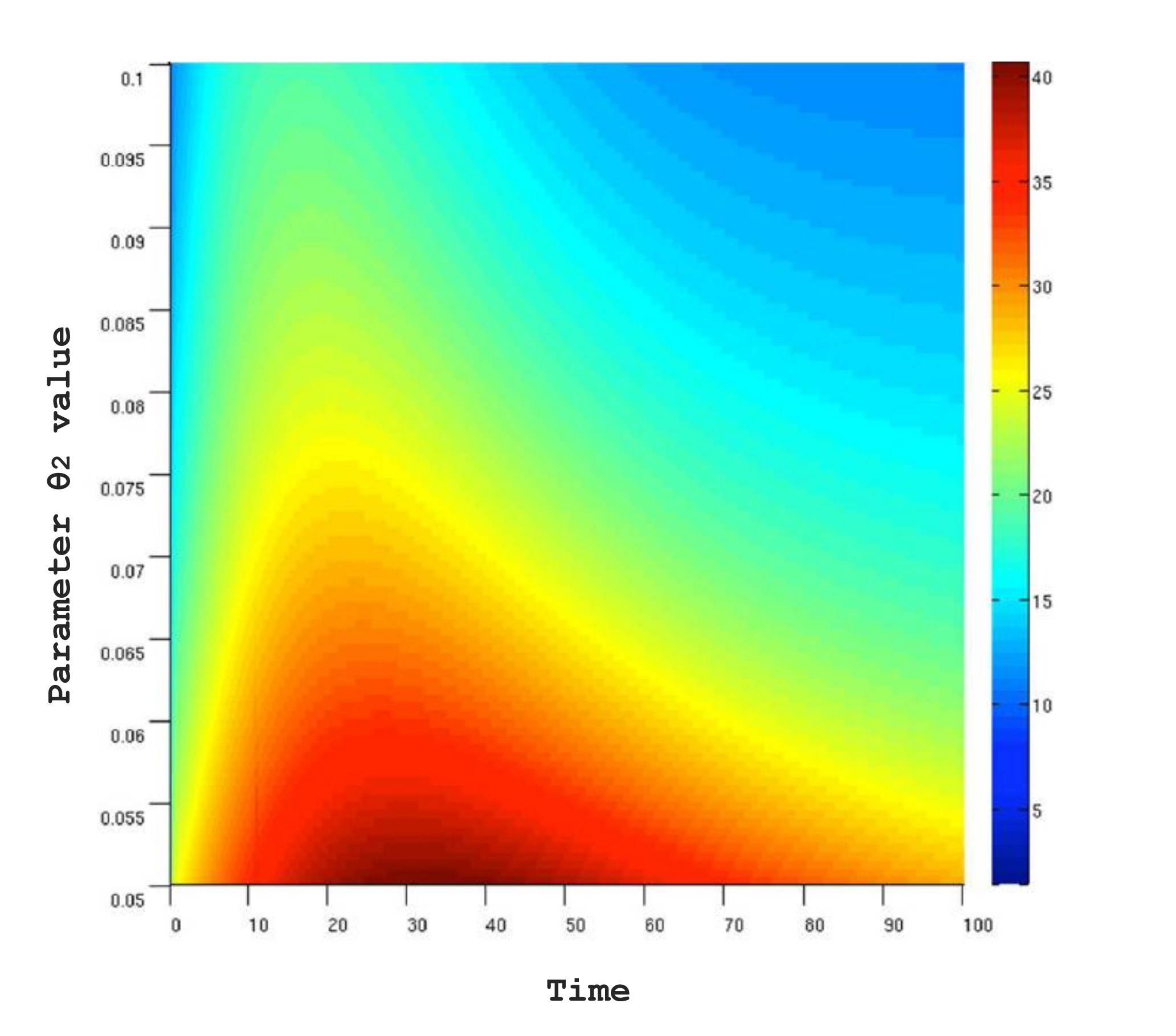}
 \caption{Approximated Fisher information of the crystallization example for different values of $\theta_2$. \label{fig:crystal_info}} 
 \vspace{-15pt}
 \end{center}
\end{floatingfigure}
\noindent Recall the simple crystallization process given by the two chemical reactions
\begin{alignat*}{2} 
& 2A  \;&\xrightarrow{\theta_1}& \;\;B,    \\
&A +C \;&\xrightarrow{\theta_2}&\;\; D.
\end{alignat*}
Let us initially assume that the value of $\theta_1 = 4$ is known and we have a prior for $\theta_2 \thicksim \mathbb{U}(0.05, 0.1)$ where $\mathbb{U}(a,b)$ refers to the continuous
uniform distribution between $a$ and $b$.
For the initial state  $(x_A, \allowbreak x_B, x_C, x_D) = (4, 0, 2 , 0)$ the number of reachable states 
is $ 7$ and no sophisticated truncation method is necessary to integrate the CME.  
Our goal is to find the best time points to take an observation for estimating $\theta_2$. 
In this case, the Fisher information is just a scalar and in Figure~\ref{fig:crystal_info} it is shown as a function of both $\theta_2$ and time. The different colors correspond to different values of $\mathcal{I}(t, \theta_2)$ (see colorbar). 
The time point for the maximum expected information is $t^* = 20.64.$ Note that in case of one unknown parameter the extension to the $R$ time points optimization problem is straightforward because of the linearity of \eqref{eq:FIM_indep}. The solution simply consists of $R$ replications of $t^*$.

\begin{figure}[tb]
\centering
\subfigure[Single observation time point.]{
 \includegraphics[width=5.7cm]{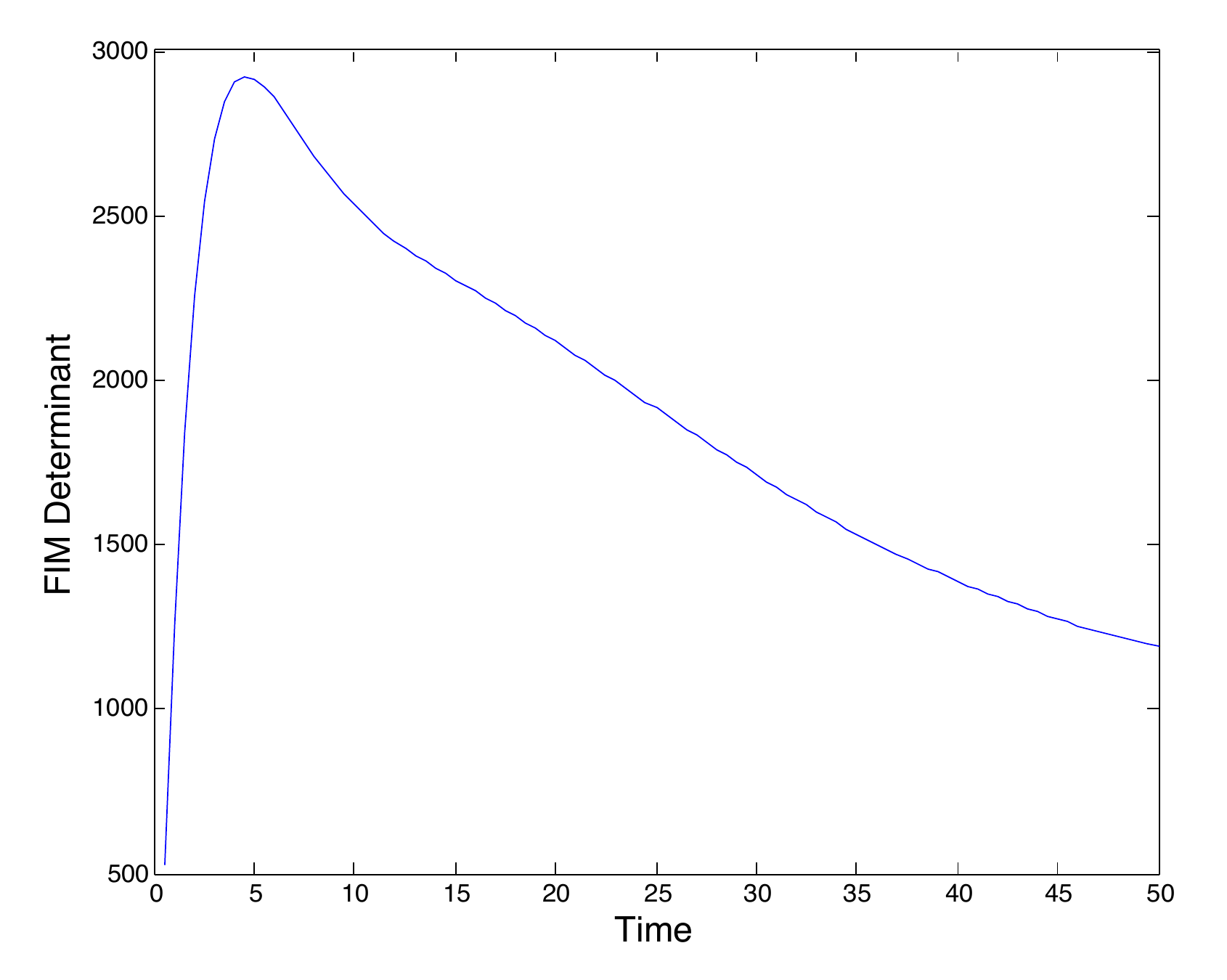} 
 \label{fig:crystal_det}
}
\subfigure[Two observation time points.]{ 
  \includegraphics[width=5.7cm]{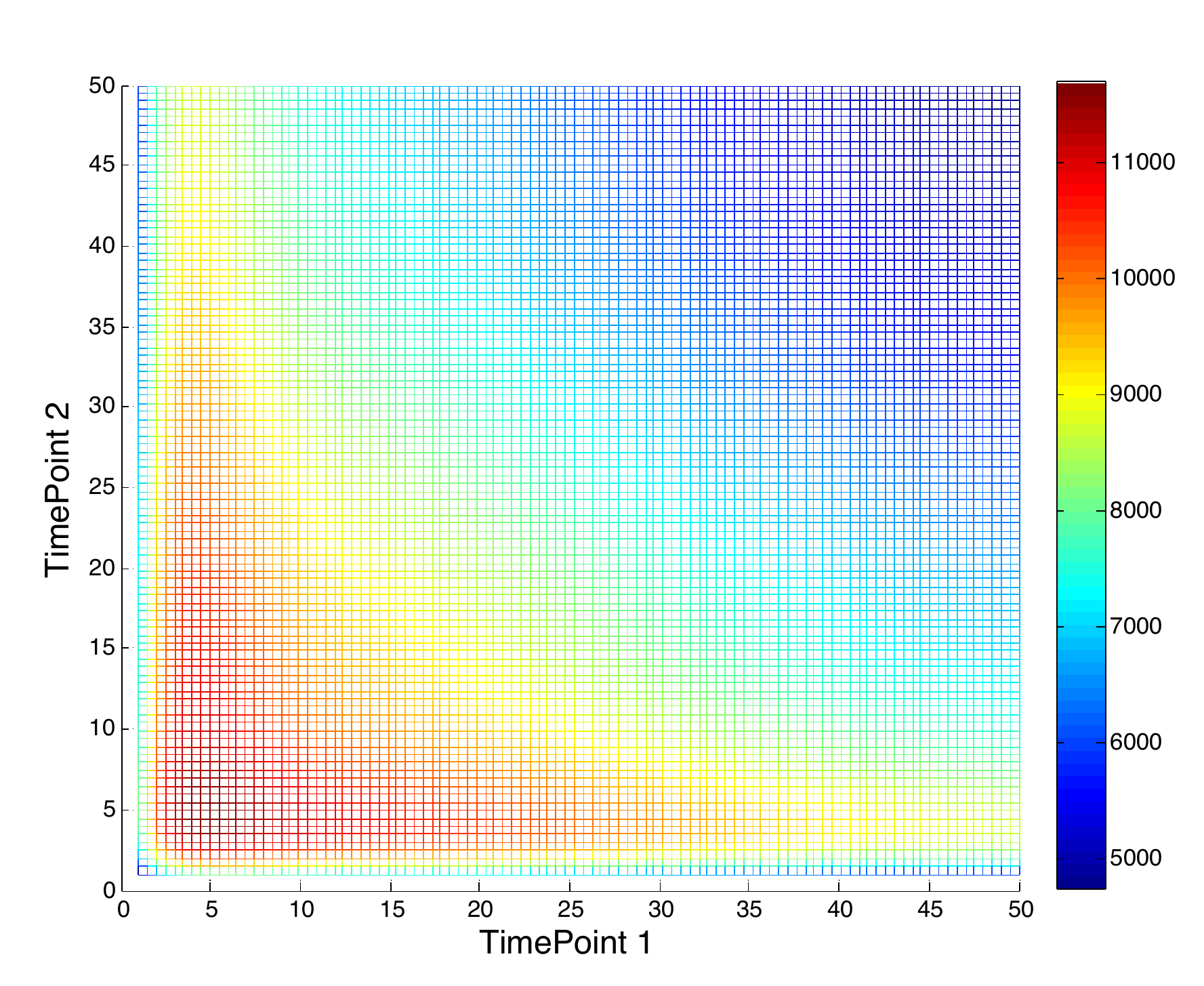} 
 \label{fig:crystal_3D}
}
\caption{Approximated determinant of the expected FIM over time for the crystallization example.}
\label{fig:test2}
\end{figure}

Next, we assume that both, $\theta_1$ and $\theta_2$ are unknown parameters
of interest. Then the Fisher information is a $2\times 2$ matrix.
 In Figure~\ref{fig:crystal_det} the evolution of the determinant of the expected FIM is shown for $\theta_1\thicksim \mathbb{U}(0.05, 0.5)$ and  $\theta_2\thicksim \mathbb{U}(0.01, 0.1).$ From the plot it is clear that if there is only one observation possible this has to be done quite early in time. Indeed, our optimal design scheme returns that the optimal time point to take an observation is at $t=4.625$ assuming that both parameters have to be estimated.

Now, we consider the case that we are able to take a second measurement $t_2 \geq t_1$.
The determinant of the average FIM is shown in Figure~\ref{fig:crystal_3D}. 
 \noindent The plot suggests that one obtains the maximum information for the estimation of both parameters 
if both     observations take place  early in the experiment. Our optimization procedure returned the optimal time points   $t_1 = t_2 = 4.625$. 
One notes that combining an early observation with a later seems also a good choice. On the other hand, taking two late observations provides significantly less information leading possibly to parameter identifiability problems.

Generalizing the above observations for the case of $R$ time points we verify from our experiments until $R = 5$ that for this particular model and this choice of parameters the optimal experiment persistently consists of taking as many as possible observations at the same early time point. We expect the same to hold for any $R \geq 2.$

\subsection{Exclusive Switch}
$~$\\[-4ex]

\begin{floatingfigure}[l]{0.4\textwidth}
\begin{center}
\vspace{-10pt}
 \includegraphics[width=0.4\textwidth]{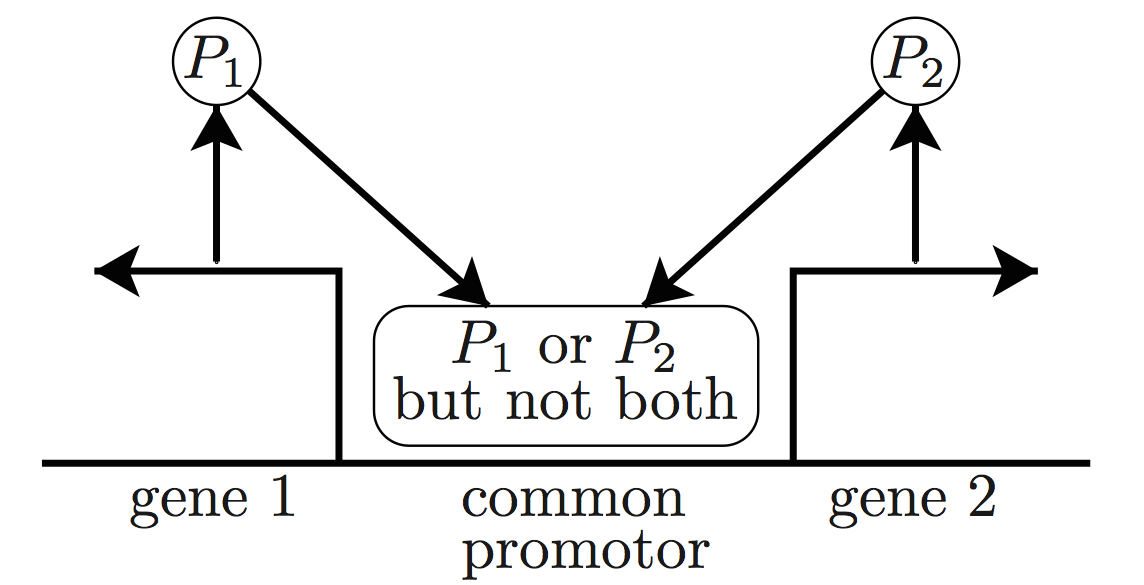}
 \caption{Exclusive Switch Network \label{fig:ex_switch}} 
 \end{center}
 \vspace{-10pt}
\end{floatingfigure}

\noindent
The exclusive switch is a gene regulatory network that consists of two genes with a common promotor region as shown in Figure \ref{fig:ex_switch}. 
The system involves five chemical species $DNA$, $P_1$, $P_2$, $DNA.P_1$, $DNA.P_2$. 
At each time point the system can be in one of the following three configurations: a) The promotor region is free, b)  $P_1$ binds to the promotor region or c) $P_2$ binds to the promotor region. 
Each of the two gene products $P_1$ and $P_2$ inhibits the expression of the other product if a molecule is bound to the promotor region. More precisely, in configuration a) (promotor region is free), molecules of both types $P_1$ and $P_2$ are produced. 
If a molecule of type $P_1 (P_2)$ is bound to the promotor region (case b) and c)), only molecules of type $P_1(P_2)$ are produced, respectively. The chemical reactions with the corresponding constant rates are shown below for  $j=\{1 ,2\}.$\\[-4ex]
\begin{alignat*}{5} 
 &DNA  &\xrightarrow{\lambda_j}& \;\; DNA + P_j \hspace{25pt}  &&\mathrm{production}\\[-1ex]
 &P_j   &\xrightarrow{\delta_j}&  \;\;\emptyset  \hspace{25pt}  &&\mathrm{degradation} \\[-1ex]
 &DNA + P_j  &\xrightarrow{\beta_j}&\;\; DNA.P_j  \hspace{25pt} && \mathrm{binding} \\[-1ex]
&DNA.P_j  &\xrightarrow{\nu_j}&\;\; DNA + P_j  \hspace{25pt}  &&\mathrm{unbinding}  \\[-1ex]
  &DNA.P_j &\xrightarrow{\lambda_j}&\;\; DNA.P_j + P_j \hspace{25pt}  &&\mathrm{bound \;production}\\[-2ex]
\end{alignat*}
 
\noindent
Depending on the chosen parameters, the probability distribution of the exclusive switch is bistable, i.e. most of the probability mass concentrates on two distinct regions of the state space. In particular, if binding to the promotor is likely, then these two regions correspond to the two configurations b) and c) where either the production of $P_1$ or the production of $P_2$ is inhibited.

For the purpose of our experiments we fixed the initial state of the system such that no proteins are present in the system and one   DNA molecule with a free promotor region. 
We set up optimal experiments for the following case:
The unknown parameters are the production and the degradation constants of $P_1$, $\lambda_1$ and $\delta_1$ respectively, while the rest of the parameter values are known. 

\begin{figure}[tb]
\centering
\subfigure[Single observation time point.]{
\includegraphics[width=5.7cm]{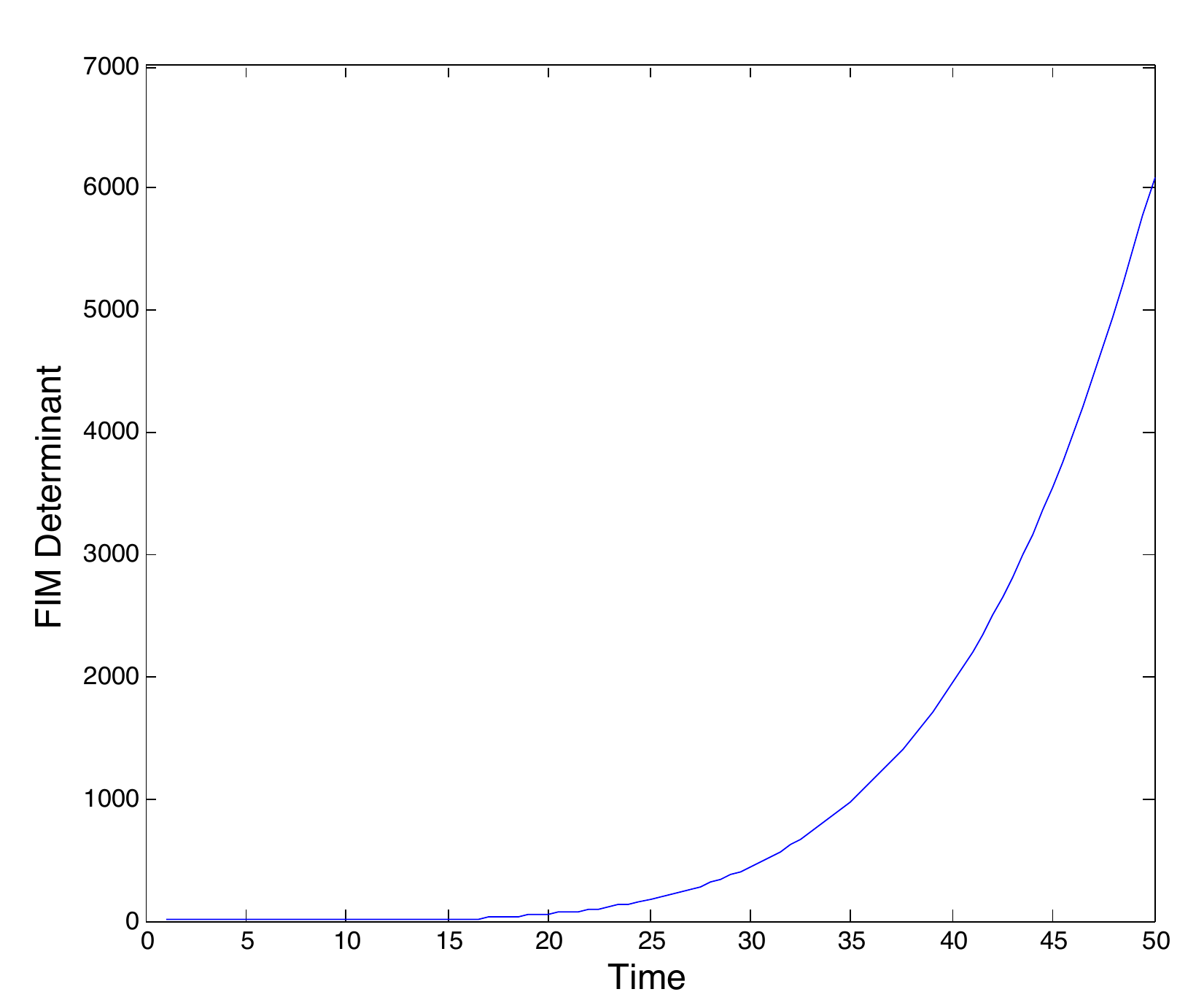}
  \label{fig:ex_switch_det}
}
\subfigure[Two observation time points.]{ 
 \includegraphics[width=5.7cm]{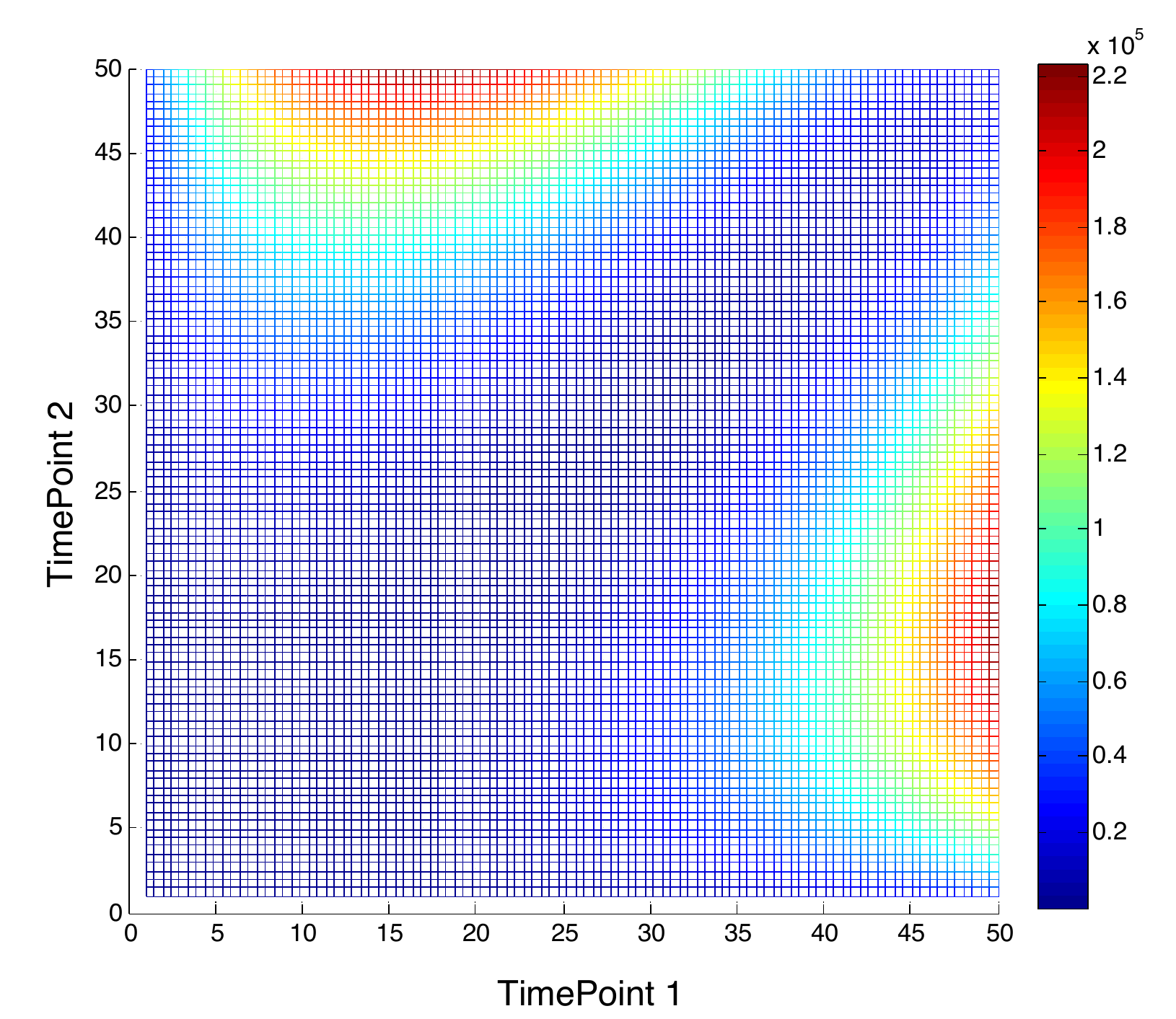}
  \label{fig:ex_switch_3D}
}
\caption{Approximated determinant of the expected FIM for the exclusive switch example (case 1).}
\label{fig:test}
\end{figure}

\noindent
We assume that for $j = \{1, 2\}$\\[-1.5ex] $$
\begin{array}{l}
\lambda_1 \thicksim\mathbb U(0.01, 0.1),
\delta_1 \thicksim \mathbb U(0.0001, 0.001), \\
\lambda_2 = 0.05, \delta_2 = 0.0005,
\beta_j = 0.001\mbox{ and }\nu_j = 0.008.\end{array}$$ 
In Figure~\ref{fig:ex_switch_det} the information is shown over time. 
From the plot it is evident that in the case of a single  observation time point one should take the measurement as late as possible in the interval $[0, 50],$ if we restrict until time $t = 50.$
This most probably arises from the fact that the chosen binding rates, $\beta_1, \beta_2$ are rather small, i.e.,   binding to the promotor is not so likely and there is a delay until the   binding influences the dynamics of  production and degradation of the corresponding proteins.

 In Figure~\ref{fig:ex_switch_3D} the information of a possible experimental setup for two time points is being presented. From the plot we can observe that the most informative experiment, now, is given by two \emph{different} time points. The second measurement is at $50$ time units, as previously, but the first one should be taken at $16.75.$ Intuitively, this could mean that for estimating multiple parameters of this model we need to observe the process at more than one time points, if possible.

Setting up optimal experiments for $R$ time points we observe, as in the first model, a replication of the optimal $p$ time points, where $p$ is the number of the unknown parameters. For $R = 3$ and $R = 4$ we get  $(t_1, t_2, t_3)= (16.75, 50, 50)$ and $(t_1, t_2, t_3, t_4) = (16.75, 16.75, 50, 50),$ respectively. 

\section{Discussion and Future Work}\label{sec:conc}
Given a stochastic model of a chemical reaction network, we computed the Fisher information of different experimental designs and determined optimal observation times. The optimality criterion that we considered was the determinant of the expected Fisher information where the expectation was taken w.r.t. some prior distribution over the unknown parameters. 

Our experimental results give rise to the conjecture that the $n$ optimal time points for a system with $p$ unknown parameters are equal replications of the $p$ optimal time points. E.g. if we have two unknown parameters, but, say, four possible observation time points, we get only two distinct times at which the Fisher information becomes maximal w.r.t. the time points.
A similar result has been proven by Box for deterministic chemical kinetics \cite{box}. We conjecture that his result carries over to the stochastic setting. This would make experiment design for stochastic chemical kinetics a much less expensive procedure since the dimension of the optimization search space reduces to the number of unknown parameters.

Other plans for future work include that we consider equidistant observation time points and optimize the time interval between two successive observations. 
We also plan to consider other ways of approximating the FIM, e.g. by using moment closure techniques and exploiting the information of a sufficiently  large number of moments.
Additionally, we will work on the case of dependent observations and develop an iterative procedure where after each optimization step experimental results become available and the next optimal observation time is computed given these results. 


\bibliographystyle{plain}
\bibliography{literature}  

\begin{thebibliography}{10}

\bibitem{CAV11}
A.~Andreychenko, L.~Mikeev, D.~Spieler, and V.~Wolf.
\newblock Parameter identification for {M}arkov models of biochemical
  reactions.
\newblock In {\em Proc.of CAV}, volume 6806 of {\em Lecture Notes of Computer
  Science}, pages 83--98. Springer, 2011.

\bibitem{MLEJournal}
A.~Andreychenko, L.~Mikeev, D.~Spieler, and V.~Wolf.
\newblock Approximate maximum likelihood estimation for stochastic chemical
  kinetics.
\newblock {\em EURASIP Journal on Bioinformatics and Systems Biology}, 9, 2012.

\bibitem{box}
G.~E.~P. Box and H.~L. Lucas.
\newblock Design of experiments in non-linear situations.
\newblock {\em Biometrika}, 46(1/2):77--90, June 1959.

\bibitem{krylovSidje}
K.~Burrage, M.~Hegland, F.~Macnamara, and B~Sidje.
\newblock A {K}rylov-based finite state projection algorithm for solving the
  chemical master equation arising in the discrete modelling of biological
  systems.
\newblock In {\em Proceedings of the {M}arkov 150th Anniversary Conference},
  pages 21--38. Boson Books, 2006.

\bibitem{gillespie77}
D.~T. Gillespie.
\newblock Exact stochastic simulation of coupled chemical reactions.
\newblock {\em J. Phys. Chem.}, 81(25):2340--2361, 1977.

\bibitem{sliding}
T.~Henzinger, M.~Mateescu, and V.~Wolf.
\newblock Sliding window abstraction for infinite {M}arkov chains.
\newblock In {\em Proc. CAV}, volume 5643 of {\em LNCS}. Springer, 2009.

\bibitem{komorowski}
M.' Komorowski, M.~J. Costa, David~A. Rand, and M.~P.~H. Stumpf.
\newblock Sensitivity, robustness, and identifiability in stochastic chemical
  kinetics models.
\newblock {\em Proceedings of the National Academy of Sciences},
  108(21):8645--8650, May 2011.

\bibitem{citeulike:821121}
L.~Ljung.
\newblock {\em {System Identification: Theory for the User (2nd Edition)}}.
\newblock {Prentice Hall PTR}, 1998.

\bibitem{loinger-lipshtat-balaban-biham07}
A.~Loinger, A.~Lipshtat, N.~Q. Balaban, and O.~Biham.
\newblock Stochastic simulations of genetic switch systems.
\newblock {\em Physical Review E}, 75:021904, 2007.

\bibitem{FAUIET}
M.~Mateescu, V.~Wolf, F.~Didier, and T.A. Henzinger.
\newblock Fast adaptive uniformisation of the chemical master equation.
\newblock {\em IET Systems Biology}, 4(6):441--452, 2010.

\bibitem{merle}
Y.~Merl\'e and F.~Mentr\'e.
\newblock Bayesian design criteria: Computation, comparison, and application to
  a pharmacokinetic and a pharmacodynamic model.
\newblock {\em Journal of Pharmacokinetics and Biopharmaceutics},
  23(1):101--125, February 1995.

\bibitem{Munsky06}
B.~Munsky and M.~Khammash.
\newblock The finite state projection algorithm for the solution of the
  chemical master equation.
\newblock {\em J. Chem. Phys.}, 124:044144, 2006.

\bibitem{pronzato1}
L.~Pronzato and E.~Walter.
\newblock Robust experiment design via stochastic approximation.
\newblock {\em Mathematical Biosciences}, 75(1):103--120, 1985.

\bibitem{Timmer}
S.~Reinker, R.M. Altman, and J.~Timmer.
\newblock Parameter estimation in stochastic biochemical reactions.
\newblock {\em IEEE Proc. Syst. Biol}, 153:168--178, 2006.

\bibitem{ruess}
J.~Ruess, A.~Milias-Argeitis, and J.~Lygeros.
\newblock Designing experiments to understand the variability in biochemical
  reaction networks.
\newblock {\em Journal of The Royal Society Interface},
  10(88):20130588--20130588, August 2013.

\bibitem{Inexact}
R.~Sidje, K.~Burrage, and S.~MacNamara.
\newblock Inexact uniformization method for computing transient distributions
  of {M}arkov chains.
\newblock {\em SIAM J. Sci. Comput.}, 29(6):2562--2580, 2007.

\bibitem{vanBos2007parameter}
A.~van~den Bos.
\newblock {\em Parameter estimation for scientists and engineers}.
\newblock Wiley-Interscience, 2007.

\end{thebibliography}

\end{document}